\documentclass[sigconf]{acmart}

\usepackage{hyperref}
\usepackage{hyperxmp}
\usepackage{xcolor}
\usepackage{tikz}
\usepackage{courier}
\usepackage{tabularx}
\usepackage{makecell}
\usepackage{booktabs}

\usepackage{xcolor}     % approved
\usepackage{framed}     % approved (and included by acmart)
\newcounter{mybox}[section]
\renewcommand{\themybox}{\thesection.\arabic{mybox}}

\definecolor{greenreference}{HTML}{45b445}

\title{Tracing Positional Bias in Financial Decision-Making: Mechanistic Insights from Qwen2.5}

\author{Fabrizio Dimino}
\affiliation{%
  \institution{Domyn}
  \city{New York}
  \country{US}
}
\email{fabrizio.dimino@domyn.com}

\author{Krati Saxena}
\affiliation{%
  \institution{Domyn}
  \city{Gurgaon}
  \country{India}
}
\email{krati.saxena@domyn.com}

\author{Bhaskarjit Sarmah}
\affiliation{%
  \institution{Domyn}
  \city{Gurgaon}
  \country{India}
}
\email{bhaskarjit.sarmah@domyn.com}

\author{Stefano Pasquali}
\affiliation{%
  \institution{Domyn}
  \city{New York}
  \country{US}
}
\email{stefano.pasquali@domyn.com}  % Title, authors, etc.

\copyrightyear{2025}
\acmYear{2025}
\setcopyright{acmlicensed}\acmConference[ICAIF '25]{6th ACM International Conference on AI in Finance}{November 15--18, 2025}{Singapore, Singapore}
\acmBooktitle{6th ACM International Conference on AI in Finance (ICAIF '25), November 15--18, 2025, Singapore, Singapore}
\acmDOI{10.1145/3768292.3770394}
\acmISBN{979-8-4007-2220-2/2025/11}

\begin{abstract}
The growing adoption of large language models (LLMs) in finance exposes high-stakes decision-making to subtle, underexamined positional biases. The complexity and opacity of modern model architectures compound this risk. We present the first unified framework and benchmark that not only detects and quantifies positional bias in binary financial decisions but also pinpoints its mechanistic origins within open-source Qwen2.5-instruct models (1.5B–14B). Our empirical analysis covers a novel, finance-authentic dataset revealing that positional bias is pervasive, scale-sensitive, and prone to resurfacing under nuanced prompt designs and investment scenarios, with recency and primacy effects revealing new vulnerabilities in risk-laden contexts. Through transparent mechanistic interpretability, we map how and where bias emerges and propagates within the models to deliver actionable, generalizable insights across prompt types and scales. By bridging domain-specific audit with model interpretability, our work provides a new methodological standard for both rigorous bias diagnosis and practical mitigation, establishing essential guidance for responsible and trustworthy deployment of LLMs in financial systems.

\noindent \textbf{Keywords}: Positional Bias, Large Language Models, Financial Decision-Making, AI Governance, Mechanistic Interpretability
\end{abstract}
\begin{document}

\maketitle

\section{Introduction}
\label{sec:introduction}

Large Language Models (LLMs) have become indispensable to modern finance, enabling workflows such as investment screening, portfolio rebalancing, and risk intelligence \citep{lopez2023finllm}. Although it is widely recognized and supported by extensive research that LLMs exhibit \textit{positional bias} (a systematic preference for alternatives based purely on their presentation order; primacy and recency bias if the model favors early or late presented options, respectively) in general open-domain scenarios \cite{shah2020predictably, wang2023large, yu2024mitigate, wang2025pine}, the financial implications of this phenomenon remain critically underexplored. In high-stakes environments where impartiality is pivotal, even subtle biases can systematically distort asset allocation, risk assessments, and regulatory compliance. Against this backdrop, the reliability of LLM-guided financial decisions warrants scrutiny, as these models transition from general-purpose applications into contexts that directly influence economic outcomes.

This work moves beyond the well-established fact of positional bias in LLMs by bringing rigorous financial context evaluation and mechanistic interpretability to the forefront. We systematically examine how positional bias manifests when open-source LLMs, specifically a popular model family, Qwen-2.5-instruct \cite{yang2024qwen2.5}, are deployed for financial decision-making through pairwise comparative assessments. Using a newly constructed dataset of finance-focused prompts that covers diverse asset classes and investor risk profiles, we not only detect and quantify positional bias, but also illuminate its inner workings with mechanistic analyses that trace bias to specific model layers and attention heads. This dual focus demonstrates, with unprecedented transparency, how general-purpose models like Qwen2.5-instruct respond to financial domain-shift, setting a foundation for targeted intervention and robust, fair decision-making in real-world financial systems. For brevity, we refer to Qwen2.5-instruct models simply as Qwen2.5 throughout the paper, unless stated otherwise.

Following are the core contributions of our paper:
\begin{enumerate}
    \item \textit{First Benchmark Integrating Positional Bias Detection and Mechanistic Interpretability in Financial LLMs}.  
        We introduce the first unified framework and dataset\footnote{On acceptance, code and the dataset will be open-sourced.} explicitly designed to benchmark both positional bias and its mechanistic origins in binary financial decision tasks, providing a comprehensive, domain-relevant standard for evaluation.
    \item \textit{Cross-Scale and Prompt-Sensitive Characterization}.  
        Our analysis systematically reveals how positional bias-both primacy and recency, varies with model size and prompt design, identifying persistent vulnerabilities unique to financial decision contexts not captured by existing general-purpose studies.
    \item \textit{Generalizable Mechanistic Insights into Bias Emergence and Propagation}.  
          Beyond traditional mapping to layers and attention heads, we deliver actionable, generalizable insights into where positional bias consistently originates in Qwen2.5 models and how it spreads across model components for different financial prompt types, enabling future paths to targeted mitigation and transparent, domain-specific model governance.
\end{enumerate}

Our paper is organized as follows. Section~\ref{sec:related_work} reviews related work. Section~\ref{sec:methods} details the methodology for detecting positional bias and interpreting it, followed by Section~\ref{sec:results}, where we report empirical findings and the insights from mechanistic interpretability. Section~\ref{sec:conclusion} concludes and outlines avenues for future research.

\section{Related Work}
\label{sec:related_work}

\paragraph{Positional Bias in General‑Purpose LLMs.}
Early studies first documented systematic positional preferences in language models through few-shot calibration experiments \cite{zhao2021calibrate}. Subsequent work confirmed sizable effects in multiple-choice reasoning \cite{wang2023large}, pairwise evaluations \cite{chen2024humans}, and binary tasks, with sensitivity to option order observed even in recent models like Qwen2.5-32B \cite{goral2024wait}. The nature and magnitude of positional bias vary substantially across model families and task domains \cite{shi2024judging, wang2024eliminating, guo2024serial}, and nuanced patterns-such as primacy bias in close-call scenarios \cite{yin2025fragile} and recency bias when quality diverges-have also been reported \cite{chen2024humans}. Cross-linguistic analyses further reveal that these effects can differ by language, challenging assumptions about uniform early-position bias for Qwen2.5‑7B \cite{menschikov2025position}. Despite this rich literature on general-purpose settings, no study has systematically mapped positional bias across different scales of Qwen2.5 models and prompt types specifically in financial pairwise tasks, nor traced its mechanistic origins in this context.

\paragraph{Bias in Financial Decision‑Making Models.}
While LLMs are increasingly integrated into financial decision-making, research on positional bias in this domain is virtually absent. Existing work focuses on demographic, social, or framing biases \cite{shah2020predictably,wang2023large} in areas like credit scoring and investment texts \cite{jin2023fairness}, or investigates behavioral patterns such as the disposition effect in LLMs cast as artificial investors \cite{tan2024dissecting}. No prior study has empirically quantified positional bias across diverse financial categories or analyzed how bias patterns shift with model scale-our work is the first to fill this gap.

\paragraph{Prompt Design, Role Assignment, and Framing.}
Prompt structure is a known driver of variability in LLM outputs: subtle changes in role assignment, framing, or constraint ordering can sharply alter model behavior, even in otherwise identical prompts \cite{hansen2024simulating,zhou2024llms}. Social bias study like \cite{hida2024social} underscore the sensitivity of models to presentation format, and recent work on Qwen2.5-7B specifically documents how reordering options within prompts shifts outcomes \cite{zeng2025order}. RAG pipeline evaluations similarly show that document ordering modulates answer accuracy in controlled settings \cite{cuconasu2025ragbias}. We further systematically examine and interpret the effect of prompt structures on positional bias in this work.

% \paragraph{Prompt Design, Role Assignment, and Framing.}
% Prompt structure has been identified as a primary driver of output variability.  Role assignment and linguistic framing can elicit dramatically different responses from the same model \cite{hansen2024simulating,zhou2024llms}. Beyond positionality itself, social‑bias studies have shown that model behavior can change sharply with subtle prompt phrasing, underscoring a high sensitivity to presentation format \cite{hida2024social}. Consistent with this, research shows that reordering constraints within otherwise identical prompts materially shifts outcomes for Qwen2.5‑7B‑Instruct \cite{zeng2025order}. Similarly, studies on RAG pipelines report that document ordering modulates answer accuracy for Qwen2.5‑7B in controlled experiments \cite{cuconasu2025ragbias}. Our empirical findings that conservative system prompts amplify positional bias, whereas moderate prompts dampen it, align with this literature and reinforce the need for prompt‑engineering safeguards in production systems.

\paragraph{Mechanistic Interpretability in Finance and Bias}
Recent work in mechanistic interpretability has uncovered how biases such as positional preferences are linked to specific architectural components-attention heads, causal masks, and positional encodings, with studies demonstrating position-sensitive hidden states \cite{yu2024mitigate}, proposing mitigation via layer-specific attention scaling \cite{adiga2024attention}, and tracing demographic and gender biases to underlying model circuits in GPT-2 and Llama2 \cite{chandna2025dissecting}. In finance, \cite{tatsat2025beyond} surveyed mechanistic interpretability’s practical roles in trading, sentiment, and hallucination detection, while \cite{gao2025evaluate} introduced BIASLENS, a vector-space method for concept-level bias analysis, and \cite{prakash2024interpreting} differentiated the roles of attention heads and MLPs in bias propagation, validating their approach on larger models. Recent mitigation strategies include PINE \cite{he2024position}, which uses bidirectional segment-level attention to eliminate positional bias, and UniBias \cite{zhou2024unibias}, which suppresses biased FFNs and attention heads at inference time. However, existing research has primarily focused on general-purpose models, particular examples or non-financial contexts, often with models such as Llama, GPT, Gemma, or Mistral, and has not yet systematically investigated how positional bias emerges, propagates, and persists in real-world financial pairwise decisions, or mapped these dynamics in open-source Qwen2.5 models. Our work directly addresses this gap by delivering the first comprehensive, model-agnostic mechanistic analysis of positional bias in Qwen2.5 across finance-specific prompt categories, rigorously tracing the internal pathways responsible for bias and providing actionable, domain-relevant transparency that prior studies have not achieved.

\section{Methodology} \label{sec:methods}

\subsection{Experimental Design}
\label{subsect:prompt_design}
Our prompt design is grounded in decision-theoretic priors. We operationalize investor risk attitudes via four advisory framings, $S$:

\begin{itemize}
    \item \textbf{Conservative}: Aligns with classical risk-aversion theory, minimizing exposure to volatility.
    \item \textbf{Moderate}: Reflects a balanced portfolio management philosophy, integrating both risk and growth opportunities.
    \item \textbf{Aggressive}: Corresponds with aggressive growth-oriented strategies, characterized by a higher tolerance for risk and greater emphasis on potential upside.
    \item \textbf{Default}: Serves as a neutral baseline without introducing intentional stylistic influence.
\end{itemize}

Recent literature underscores the pronounced responsiveness of language models to additional authoritative input \cite{anagnostidis2024susceptible}. Thus, our system prompts follow a structured directive format: \emph{"You are a [conservative/moderate/aggressive] investment advisor. When presented with two company ticker symbols, you must select exactly one based on specified criteria."}

To span common lenses of equity analysis, we define a task taxonomy of ten prompt categories $U$, motivated by evidence that template design shapes LLM behavior \cite{wei2024systematic}:

\begin{itemize}
\item \textit{Fundamental}: Evaluating financial health.
\item \textit{Sentiment}: Incorporating market sentiment.
\item \textit{ESG}: Analyzing adherence to ESG criteria.
\item \textit{Technical}: Assessing stocks performance indicators.
\item \textit{Risk}: Focusing explicitly on risk management practices.
\item \textit{Growth}: Investigating future growth potential and trajectory.
\item \textit{Dividend}: Prioritizing dividend yield consistency.
\item \textit{Sector Leadership}: Identifying industry competitive advantages.
\item \textit{Innovation}: Highlighting investments in R\&D.
\item \textit{Generic}: Offering a general investment attractiveness.
\end{itemize}

Considering prior evidence of the impact of prompt structure on model performance \cite{mao2023prompt}, we implement two distinct prompt template to control for ordering effects and grammatical bias:
\begin{itemize}
    \item Prompt 1/Template 1: \emph{"Between {company1} and {company2}, which is the better investment based on [fundamental/ESG/etc.]? Answer with only the ticker."}
    \item Prompt 2/Template 2: \emph{"Based on [fundamental/ESG/etc.], which is the better investment: {company1} or {company2}? Answer with only the ticker symbol."}
\end{itemize}

This dual-structure approach addresses potential primacy effects while maintaining response consistency through grammar-constrained outputs, a methodology validated in recent computational linguistics research\cite{wang2023grammar,geng2023grammar}.

\paragraph{Universe and sampling}
% We analyze a universe of $N=18$ large–cap U.S. technology and media firms commonly referred to as \textit{FAANG+}. This acronym extends the original FAANG grouping-\textbf{F}acebook (Meta), \textbf{A}pple, \textbf{A}mazon, \textbf{N}etflix, and \textbf{G}oogle (Alphabet)-by including additional firms with comparable market influence and technological relevance. In our case, the extended set includes companies such as Microsoft, Nvidia, Tesla, and others, yielding a total of 18 firms. These firms are selected due to their high visibility in financial markets, strong analyst coverage, and frequent appearance in institutional portfolios, making them a representative benchmark for LLM-based financial evaluation.
We analyze $N=18$ large-cap U.S. technology and media firms—an extended \textit{FAANG+} set including Microsoft, Nvidia, Tesla, and others. These companies are highly visible in financial markets, heavily covered by analysts, and commonly held in institutional portfolios, making them a representative benchmark for LLM-based financial evaluation.

All ordered pairs of distinct firms are considered: $P = N(N-1) = 18 \times 17 = 306$. Each pair is evaluated under $U=10$ prompt categories, $O=2$ prompt orderings, and $S=4$ system styles, for a total of $V = U \times O \times S = 80$ configurations per pair. To control for stochastic variability, every configuration is replicated $R=3$ times. Hence, the total number of observations is $|\mathcal{D}| = P \times V \times R = 306 \times 80 \times 3 = \boxed{73{,}440}.$

\subsection{Positional Bias Detection}

To detect positional bias systematically, we implement a paired comparison experimental design. For each ordered company pair $(i,j)$, prompt category $c$, ordering $o \in \{1,2\}$ (1 = \(i\) first, 2 = \(j\) first), and replication $r \in \{1,\dots,n\}$, the model decision function
\begin{equation}
    f_c(i,j,o,r) \mapsto \big(\hat{y}_{ijcor},\, p_{ijcor}(\hat{y}_{ijcor})\big)
\end{equation}

returns the selected company $\hat{y}_{ijcor} \in \{i,j\}$ and an associated probability $p_{ijcor}(\hat{y}_{ijcor}) \in [0,1]$. Let $\ell_{ijcor}(i)$ and $\ell_{ijcor}(j)$ denote the corresponding logits. Probabilities are obtained via the softmax:
\begin{equation}
    p_{ijcor}(x) \;=\; \frac{\exp\{\ell_{ijcor}(x)\}}{\exp\{\ell_{ijcor}(i)\} + \exp\{\ell_{ijcor}(j)\}}, \qquad x \in \{i,j\}.
\end{equation}

Given inherent stochastic variability in LLM outputs, each comparison is independently replicated 3 times. This replication strategy is chosen to balance computational efficiency and statistical reliability, aligning with best practices for evaluating consistency in LLM outputs \cite{zhao2021calibrate,wang2023large,shi2024judging,wei2024systematic}. To quantify positional effects for firm $i$ in category $c$, we average probabilities across replications for each ordering,
\begin{equation}
\bar{p}_{i,c}^{(1)} \;=\; \frac{1}{n}\sum_{r=1}^{n} p_{ijc1r}(i), 
\qquad
\bar{p}_{i,c}^{(2)} \;=\; \frac{1}{n}\sum_{r=1}^{n} p_{jic2r}(i),
\end{equation}
and define the positional difference
\begin{equation}
\Delta_{i,c} \;=\; \bar{p}_{i,c}^{(1)} - \bar{p}_{i,c}^{(2)}.
\end{equation}
A positive (negative) $\Delta_{i,c}$ indicates that firm $i$ is assigned higher (lower) probability when presented first rather than second. 

\subsection{Statistical Inference}

To robustly analyze positional bias, we adopt an \emph{estimation-first} statistical inference framework. This method integrates a non-parametric hypothesis test with a rigorous, uncertainty-aware effect size analysis, structured around three key components: (i) a two-sided Wilcoxon signed-rank test, (ii) a standardized effect sizes, and (iii) a cluster-bootstrap confidence intervals.

Let $\Delta = (d_1,\dots,d_N)$ denote the vector of paired differences. We first apply a two-tailed Wilcoxon signed-rank test

\[
H_0 : Q_{0.5}(\Delta) = 0,
\qquad
H_1 : Q_{0.5}(\Delta) \neq 0.
\]

at significance level $\alpha = 0.05$. To quantify the magnitude of the observed effect, we report the effect size statistic $r$, computed as:

\begin{equation}
    r = \frac{Z}{\sqrt{N}}.
\end{equation}

where $Z$ denotes the standardized test statistic obtained from the Wilcoxon test, and $N$ represents the number of non-zero paired differences.

To estimate location shift and uncertainty, we compute the \emph{Hodges– Lehmann} estimator  

\begin{equation}
   \hat\theta_{\mathrm{HL}}=Q_{0.5}\{(d_i+d_j)/2\} 
\end{equation}

a robust $U$‑statistic with superior small‑sample efficiency \cite{hodges1963estimation}. Sampling variability is quantified by a percentile \textbf{cluster bootstrap} ($B=5{,}000$ resamples). The 95\,\% confidence interval is

\begin{equation}
   \text{CI}_{95\%}= \bigl[\hat\theta_{\mathrm{HL}}^{(2.5\%)},\;\hat\theta_{\mathrm{HL}}^{(97.5\%)}\bigr]. 
\end{equation}

% \usepackage{graphicx}
\begin{table*}[!h]
\caption{Empirical Results for Model Scale (Prompt 1, Default) and Prompt Ordering (Default, 14B)}
\label{tab:model_size_and_ordering}
\resizebox{\textwidth}{!}{%
\begin{tabular}{l|llllll|llll}
\hline
                   & \multicolumn{6}{c|}{Model Scale}                                                                                                                  & \multicolumn{4}{c}{Prompt Ordering}                                                             \\ \cline{2-11} 
                   & \multicolumn{2}{c}{1.5B}                       & \multicolumn{2}{c}{7B}                         & \multicolumn{2}{c|}{14B}                        & \multicolumn{2}{c}{Prompt1}                    & \multicolumn{2}{c}{Prompt2}                    \\ \cline{2-11} 
Category           & \multicolumn{1}{c}{r} & \multicolumn{1}{c}{HL} & \multicolumn{1}{c}{r} & \multicolumn{1}{c}{HL} & \multicolumn{1}{c}{r} & \multicolumn{1}{c|}{HL} & \multicolumn{1}{c}{r} & \multicolumn{1}{c}{HL} & \multicolumn{1}{c}{r} & \multicolumn{1}{c}{HL} \\ \hline
ESG                & 0.87***               & 76.5 {[}70.9,81.2{]}   & 0.86***               & 49.3 {[}47.6,50.8{]}   & \bfseries 0.43***               & –2.0 {[}–6.0,–0.3{]}    & \bfseries 0.43***               & -2.0 {[}-6.0,-0.3{]}   & 0.48***               & -2.5 {[}-9.5,-0.4{]}   \\
Sentiment          & 0.87***               & 82.7 {[}77.4,90.1{]}   & 0.83***               & 35.2 {[}27.9,39.3{]}   & \bfseries 0.62***               & –2.1 {[}–9.1,–0.5{]}    & \bfseries 0.62***               & -2.1 {[}-9.1,-0.5{]}   & 0.70***               & -1.4 {[}-16.0,-0.1{]}  \\
Technical          & 0.87***               & 82.7 {[}77.4,90.1{]}   & 0.87***               & 54.0 {[}50.0,60.4{]}   & \bfseries 0.24**                & 0.0 {[}–0.4,0.0{]}      & \bfseries 0.24**                & 0.0 {[}-0.4,0.0{]}     & 0.28**                & 0.0 {[}-0.1,0.0{]}     \\
Fundamental        & 0.87***               & 81.4 {[}75.1,85.9{]}   & 0.87***               & 82.7 {[}74.0,90.0{]}   & \bfseries 0.62***               & 7.9 {[}2.2,27.8{]}      &  0.62***               & 7.9 {[}2.2,27.8{]}     & \bfseries 0.59***               & 1.6 {[}0.2,9.1{]}      \\
Growth             & 0.87***               & 74.6 {[}67.1,81.0{]}   & 0.87***               & 36.5 {[}25.4,44.0{]}   & \bfseries 0.53***               & 0.0 {[}–0.1,0.0{]}      & \bfseries 0.53***               & 0.0 {[}-0.1,0.0{]}     & 0.75***               & -0.2 {[}-1.1,0.0{]}    \\
Innovation         & 0.86***               & 64.9 {[}59.4,69.6{]}   & 0.83***               & 23.1 {[}16.0,33.2{]}   & \bfseries 0.17*                 & 0.0 {[}0.0,0.0{]}       & \bfseries 0.17*                 & 0.0 {[}0.0,0.0{]}      & 0.72***               & -0.1 {[}-0.6,0.0{]}    \\
Risk               & 0.87***               & 62.1 {[}57.1,68.1{]}   & 0.82***               & 31.9 {[}24.2,36.4{]}   & \bfseries 0.76***               & 28.1 {[}18.2,34.9{]}    & \bfseries 0.76***               & 28.1 {[}18.2,34.9{]}   & 0.77***               & 6.3 {[}0.9,34.0{]}     \\
Dividend           & 0.87***               & 68.1 {[}63.0,73.1{]}   & 0.87***               & 59.0 {[}53.9,68.0{]}   & \bfseries 0.11ns                & 0.0 {[}0.0,0.3{]}       & \bfseries 0.11ns                & 0.0 {[}0.0,0.3{]}      & 0.26**                & 0.0 {[}0.0,0.0{]}      \\
Sector\_Leadership & 0.87***               & 75.2 {[}68.4,80.1{]}   & 0.87***               & 46.7 {[}42.6,48.9{]}   & \bfseries 0.20**                & 0.0 {[}–0.1,0.0{]}      & \bfseries 0.20**                & 0.0 {[}-0.1,0.0{]}     & 0.29***               & 0.0 {[}-0.4,0.0{]}     \\
Generic            & 0.87***               & 72.5 {[}61.7,79.5{]}   & 0.86***               & 50.8 {[}49.3,57.5{]}   & \bfseries 0.23**                & 0.0 {[}–0.2,0.0{]}      & \bfseries 0.23**                & 0.0 {[}-0.2,0.0{]}     & 0.37***               & 0.0 {[}-0.3,0.0{]}     \\ \hline
\end{tabular}%
}
% \begin{tablenotes}[flushleft]\footnotesize
%       \item $^{***}p<.001$, $^{**}p<.01$, $^{*}p<.05$, $^{ns}p\ge .05$
%     \end{tablenotes}
\caption*{\footnotesize \textit{Notes.} $^{***}p<.001$, $^{**}p<.01$, $^{*}p<.05$, $^{ns}p\ge .05$.}
\end{table*}

\subsection{Interpreting Positional Bias}
\label{subsect:method_interp}
To provide mechanistic insights into positional bias, we employ the TransformerLens library\footnote{\url{https://github.com/TransformerLensOrg/TransformerLens}} to instrument Qwen2.5 models. This setup enables layer and position-specific access to model activations and outputs, facilitating a series of attribution analyses tied directly to the key prediction outcome: the model’s preference for one company ticker ($t$) over another, operationalized as the logit difference between target subword tokens (i.e., $\mathrm{logit}(t_1) - \mathrm{logit}(t_2)$).

\noindent \textbf{Direct Logit Attribution (DLA): } The Direct Logit Attribution (DLA) approach assesses how each layer and token position contributes to the comparative preference between the two entities. For each prompt, we identify the first subword tokens $t_1$ and $t_2$ for the two company tickers and capture the residual activation vector $r_{p}^{(l)}$ at every layer $l$ and token position $p$. By projecting this residual through the unembedding matrix $W_U$, we obtain logits:
\begin{equation}
    L_p^{(l)} = W_U(r_{p}^{(l)})
\end{equation}
We then define the attribution score as:
\begin{equation}
    A_p^{(l)} = L_p^{(l)}[t_1] - L_p^{(l)}[t_2]
\end{equation}
Iterating over all layers and positions, we collect tuples $(l, p, A_p^{(l)})$ for each prompt, revealing where positional bias manifests most strongly.

\noindent \textbf{Logit Lens Ranking Analysis: } To complement DLA, we use a logit lens ranking strategy at the final token position (typically where the model’s preference is expressed). For each layer $l$, we extract the final-position residual $r_{p}^{(l)}$, compute its projection, and determine the ranks 
$r_{1}^{(l)}$, $r_{2}^{(l)}$ of $t_1$ and $t_2$ after sorting all logits in descending order. The rank-difference metric is then defined:
\begin{equation}
    \Delta r^{(l)} = r_{1}^{(l)} - r_{2}^{(l)}
\end{equation}
Aggregating this across all prompts, we compute the win-rate $W[\Delta r^{(l)} < 0]$ (frequency with which $t_2$ outranks $t_1$), the mean and the median rank difference per layer. This provides a layer-wise summary of bias directionality and consistency, crucial for pinpointing where comparative biases are introduced or amplified.

\noindent \textbf{Attention Head Ablation Attribution: } To dissect which attention heads are most responsible for positional bias, we employ head-ablation within the TransformerLens framework. For each prompt, we compute ``baseline" (original) logits and their residuals, then systematically zero out individual attention heads at every layer, re-running the model to obtain ``ablated" outputs. The head-level attribution score is:
\begin{equation}
    A^{(l, h)} = (L_{base}[t_1] - L_{base}[t_2]) - (L_{abl}^{(l,h)}[t_1] - L_{abl}^{(l,h)}[t_2])
\end{equation}
This decomposition makes it possible to locate systematic, prompt-sensitive origins of bias in the attention heads.
\section{Experimental Results} 
\label{sec:results}

\begin{table*}[]
\centering
\caption{Empirical Results for Prompt System Style (14B, Prompt 1)}
\label{tab:prompt_styles_14B}
\resizebox{0.6\textwidth}{!}{%
\begin{tabular}{l|llllll}
\hline
                   & \multicolumn{6}{c}{Prompt System Style}                                                                                                           \\ \cline{2-7} 
                   & \multicolumn{2}{c}{Aggressive}                 & \multicolumn{2}{c}{Moderate}                   & \multicolumn{2}{c}{Conservative}                \\ \cline{2-7} 
Category           & \multicolumn{1}{c}{r} & \multicolumn{1}{c}{HL} & \multicolumn{1}{c}{r} & \multicolumn{1}{c}{HL} & \multicolumn{1}{c}{r} & \multicolumn{1}{c}{HL}  \\ \hline
ESG                & 0.43***               & -4.4 {[}-11.0,-0.4{]}  & \bfseries 0.41***               & -2.4 {[}-7.4,-0.9{]}   & 0.52***               & -1.3 {[}-4.8,-0.4{]}    \\
Sentiment          & 0.64***               & -1.2 {[}-2.4,-0.2{]}   & 0.60***               & -3.8 {[}-11.0,-1.0{]}  & \bfseries 0.18*                 & 0.0 {[}0.0,0.0{]}       \\
Technical          & 0.28***               & 0.0 {[}-0.1,0.0{]}     & 0.29***               & -0.1 {[}-0.4,-0.0{]}   & 0.24**                & 0.0 {[}0.0,0.7{]}       \\
Fundamental        & 0.47***               & 1.4 {[}0.3,6.2{]}      & 0.67***               & 5.9 {[}3.6,13.4{]}     & 0.78***               & 0.3 {[}0.1,1.2{]}       \\
Growth             & 0.41***               & 0.0 {[}-0.0,0.0{]}     & 0.63***               & -0.1 {[}-0.6,-0.0{]}   & 0.85***               & -48.8 {[}-49.8,-45.6{]} \\
Innovation         & 0.05ns                & 0.0 {[}0.0,0.0{]}      & 0.01ns               & 0.0 {[}0.0,0.0{]}      & 0.17*                 & 0.1 {[}0.0,0.4{]}       \\
Risk               & 0.52***               & 3.7 {[}1.7,7.4{]}      & 0.78***               & 24.7 {[}5.4,30.1{]}    & 0.80***               & 0.0 {[}0.0,0.5{]}       \\
Dividend           & 0.17*                 & 0.2 {[}0.0,2.0{]}      & 0.17*                 & 0.0 {[}0.0,0.4{]}      & 0.48***               & 0.0 {[}0.0,0.0{]}       \\
Sector\_Leadership & 0.00ns                & 0.0 {[}0.0,0.0{]}      & 0.04ns                & 0.0 {[}-0.0,0.0{]}     & 0.69***               & 0.3 {[}0.0,1.9{]}       \\
Generic            & 0.33***               & 0.0 {[}-0.0,0.0{]}     & 0.09ns                & 0.0 {[}-0.1,0.0{]}     & 0.73***               & 0.0 {[}0.0,0.2{]}     
\\ \hline
\end{tabular}%

}
\end{table*}

To systematically interrogate the depth, variability, and origins of positional bias in open-source Qwen2.5 models under finance-domain binary comparative prompts, we structure our empirical investigation around the following six core research questions:
\begin{enumerate}
    \item \textbf{RQ1:} Does increasing model scale reliably attenuate positional bias across financial categories?
    \item \textbf{RQ2:} How sensitive is positional bias to prompt ordering under a fixed system style, and which categories exhibit the largest order‑induced shifts in effect size $r$ and Hodges–Lehmann (HL) estimators?
    \item \textbf{RQ3:} How do system framing styles interact with model scale to amplify or dampen positional bias?
    \item \textbf{RQ4:} Are there consistent internal components that drive positional bias across diverse financial prompt categories and model sizes?
    \item \textbf{RQ5:} How do category semantics and template wording influence the magnitude and propagation of positional bias in Qwen2.5 across different model scales?
    \item \textbf{RQ6:} What shared or divergent mechanistic features emerge across model scales and prompt types, and how can these insights inform universal versus context-specific de-biasing strategies for financial LLMs?
\end{enumerate}
We conduct the experiments on the open-source Qwen2.5 model family (1.5B, 7B, and 14B parameters) and we evaluate all the models in a zero-shot setting.

\begin{figure*}[h]
    \centering
    \includegraphics[width=0.8\textwidth]{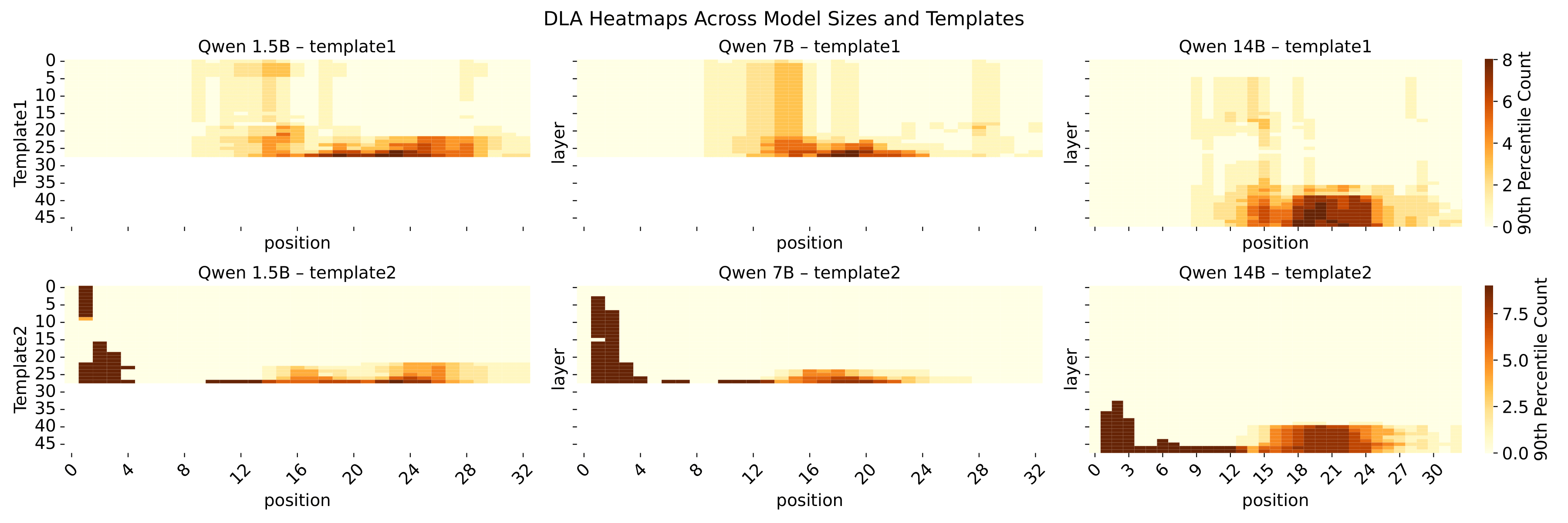} 
    \caption{DLA results on Layer/Position generalized across all categories for each model scale and prompt template.}
    \label{fig:rq4_layerpos}
\end{figure*} 

\subsection{RQ1-RQ3: Empirical Results and Insights}
\label{subsect:rq3rq5}
\textbf{RQ1: Does increasing model scale reliably attenuate positional bias across financial categories?}
% Building upon prior findings \cite{gupta2023robust}, which indicate that larger models generally demonstrate more sophisticated behavior and reduced bias, our analysis evaluated how variations in model scale influence positional-prompt bias. Specifically, we compute effect sizes ($r$) and Hodges–Lehmann (HL) estimators to uncover systematic patterns in how this bias evolves with increasing model capacity. To isolate the effect of model scale from other confounding factors, this analysis was conducted using Prompt 1 with the Default prompt style held constant across all model variants.
Building on prior work \cite{gupta2023robust} showing larger models exhibit less bias, we analyze how model scale affects positional-prompt bias using effect sizes ($r$) and Hodges- Lehmann estimators. Focusing solely on Prompt 1 with a fixed Default style, this isolates scale effects to reveal systematic bias evolution as model capacity increases.

The results in \autoref{tab:model_size_and_ordering}[left] illustrate a clear decrease in positional bias with increased model scale. The 1.5B and 7B models consistently exhibit strong positional biases across all categories, whereas the 14B model shows significantly diminished bias in most categories. Notably, \textit{ESG} and \textit{Sentiment} are the only categories exhibiting reversed bias (negative HL estimators), suggesting that, in larger models, positional preferences may invert, with a potential shift toward favoring the second-listed option.

However, an important caveat to this trend was observed in the \textit{Risk} category. Despite increased model complexity, the 14B model continued to exhibit a high level of positional bias, suggesting that the semantic ambiguity of terms such as “volatility” and “exposure” may induce an intrinsic positional bias that is resistant to scaling. Thus, it is critical to recognize that bias mitigation through model scaling might vary significantly across different reasoning domains.

\textbf{RQ2: How sensitive is positional bias to prompt ordering under a fixed system style, and which categories exhibit the largest order‑induced shifts in effect size $r$ and Hodges–Lehmann (HL) estimators?}
To further explore the sensitivity of positional bias to variations in prompt ordering, we analyzed outcomes within the 14B model with the Default prompt style using two distinct prompt arrangements.

The results in \autoref{tab:model_size_and_ordering}[right] demonstrated that even subtle changes in prompt sequencing significantly influenced effect sizes, indicating that prompt design plays a critical role in either exacerbating or mitigating positional bias.

Categories such as \textit{Growth} and \textit{Innovation} displayed pronounced differences between the two prompts, suggesting these categories may be especially sensitive to ordering effects. Conversely, the \textit{Fundamental} and \textit{Risk} categories exhibited relatively consistent bias across prompt variations, reinforcing the notion that positional biases may be inherently tied to the nature of the decision-making domain rather than solely to the presentation order.

\textbf{RQ3: How do system framing styles interact with model scale to amplify or dampen positional bias?}
Recognizing that specific prompt styles can potentially influence model behavior and bias \cite{hida2024social, guo2024serial}, we evaluated the effect of different prompt framing styles on positional bias using the 14B model with Prompt 1.

\autoref{tab:prompt_styles_14B} highlights that prompt styles play a substantial role in bias modulation. The \textbf{Aggressive} prompt style typically resulted in intermediate effect sizes, reflecting a balanced but growth-oriented decision-making approach. In contrast, the \textbf{Conservative} framing systematically amplifies positional bias in most of the categories, plausibly because its cautious, directive language reinforces primacy heuristics. Interestingly, the \textbf{Moderate} style often exhibited the most nuanced results, with lower positional biases in several categories. This suggests that balanced framing may facilitate more robust reasoning by limiting overly simplistic heuristic biases.

Taken together, these results indicate that prompt style is not merely a cosmetic choice but a substantive methodological lever for bias control. Conservative, risk‑averse framings tend to amplify positional dependencies (with a few systematic exceptions such as \textit{Sentiment} and \textit{Technical}); moderate framings attenuate them in several economically salient categories (e.g., \textit{ESG}, \textit{Generic}); and aggressive framings can nearly abolish them in targeted domains (e.g., \textit{Sector Leadership}, \textit{Innovation}).

\begin{figure*}[h]
    \centering
    \includegraphics[width=0.8\textwidth]{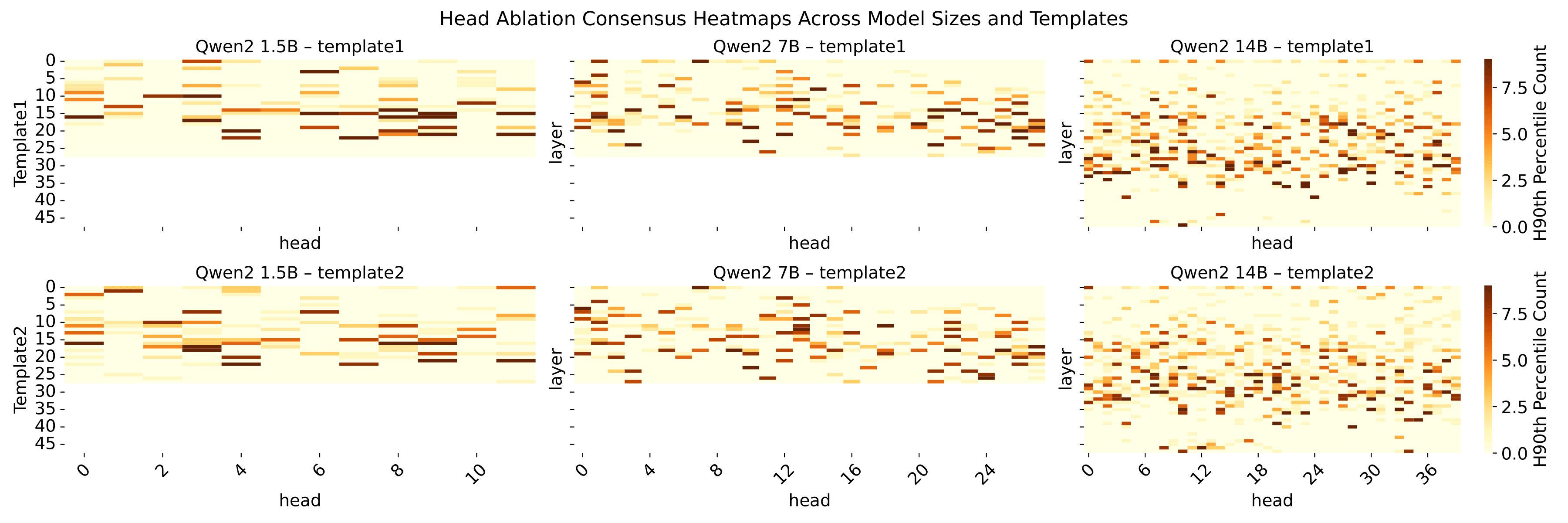} 
    \caption{Head Ablation results generalized across all categories for each model scale and prompt template.}
    \label{fig:rq4_head}
\end{figure*}

\begin{figure*}[h]
    \centering
    \includegraphics[width=0.8\textwidth]{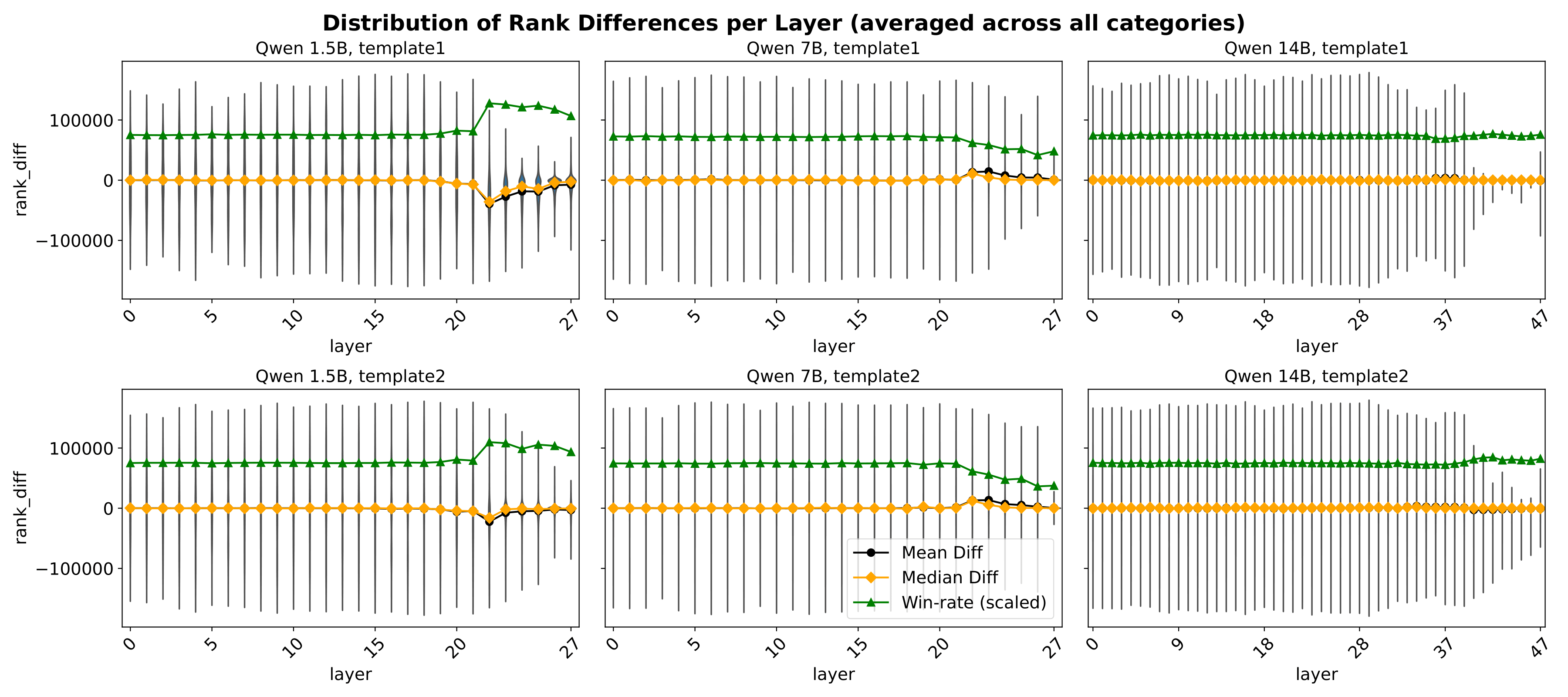} % Replace with your image file
    \caption{Rank distribution per layer and Win-rates across all categories for each model scale and template.}
    \label{fig:rq5}
\end{figure*}
\subsection{RQ4-RQ6: Mechanistic Interpretability Insights}
\label{subsect:rq3rq5}
\noindent \textbf{RQ4: Are there consistent internal components that drive positional bias across diverse financial prompt categories and model sizes?}
To determine consistent internal components driving positional bias, we aggregated Direct Logit Attribution (DLA) and head ablation results across all categories and both prompt templates. We constructed consensus plots by selecting layers, token positions, and attention heads exceeding the 90th percentile attribution thresholds, highlighting the strongest, most frequent bias sources irrespective of prompt phrasing or task specifics.

Figure \ref{fig:rq4_layerpos} show a robust pattern: positional bias concentrates in the mid-to-late transformer layers, typically layers 12 to 24 in 1.5B and 7B model and layer 32 to 48 in 14B model, and notably affects token positions corresponding to the second half of prompts-where options are presented. This holds true across both prompt templates, with Template 2 also showing early-layer bias due to its frontloading of options. These results underscore that, while prompt framing shifts the timing of bias introduction (early vs. middle positions), the main locus of bias consistently emerges in deeper layers responsible for integrating and comparing options. Complementing this, Figure \ref{fig:rq4_head} identifies a small but stable subset of attention heads within these layers driving bias. Smaller models feature fewer of these “super-biased” heads, while larger models distribute bias more widely but still share a core set. Across templates, we found the overlap of 39.5\%, 45.7\% and 41.8\% of these heads across the two templates in Qwen 1.5B, 7B and 14B models, respectively. These results establish that, across scales, prompts, and financial contexts, mid‑to‑late transformer layers are the universal locus. A small, overlapping set of attention heads within those layers act as the bias “engines.” Prompt framing tweaks when the bias injects (early vs. late) but not which layers/heads ultimately carry it.

\begin{figure*}[h]
    \centering
    \includegraphics[width=0.8\textwidth]{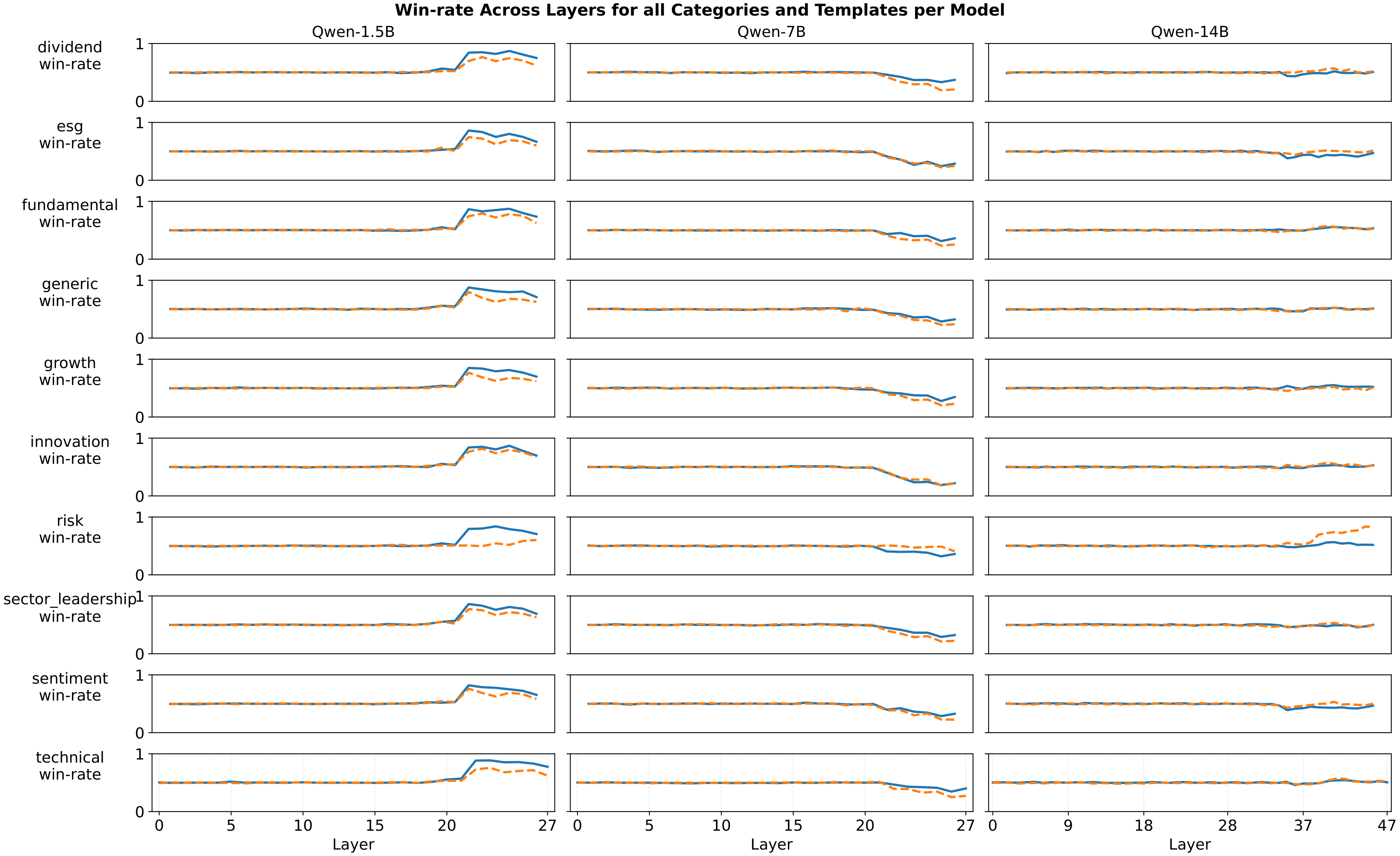} % Replace with your image file
    \caption{Win-rates for all categories, model scales and templates.}
    \label{fig:rq5_individual}
\end{figure*}

\noindent \textbf{RQ5: How do category semantics and template wording influence the magnitude and propagation of positional bias in Qwen2.5 across different model scales?}
We analyze positional bias magnitude and its propagation across financial prompt categories, template designs, and model scales using rank difference metrics and win-rate statistics. These results illuminate where bias emerges and intensifies through transformer layers, linking semantic content and prompt phrasing to internal bias dynamics within Qwen2.5 models.

As shown in Figure \ref{fig:rq5}, across all three model sizes (1.5B, 7B, and 14B), positional bias is minimal in early layers, with mean and median rank differences near zero, indicating little initial preference. Bias sharply intensifies in mid-to-late layers-typically after layer 20 in smaller models and around layers 35–40 in the largest model-reflecting deeper layers as integration points for comparative judgments. Smaller models exhibit wider spreads and more abrupt changes, signaling more pronounced and less calibrated biases, with 1.5B favoring the first position and 7B/14B slightly favoring the second, based on where the win-rate plot moves. As illustrated in Figure \ref{fig:rq5_individual}, most categories follow similar bias patterns, except the \textit{Risk} category, which shows pronounced spikes especially in smaller models. Template 2 tends to reduce bias variation for risk in smaller models, suggesting prompt phrasing can moderate bias more effectively at lower scales. In 7B, \textit{Innovation}, \textit{Sector Leadership}, and \textit{Growth} categories show deeper win-rate dips, indicating stronger bias fluctuations influenced by semantic richness and prompt design.

Prompt wording modulates but does not fundamentally change bias trends: Template 2 (options before criteria) slightly dampens bias spikes, particularly in smaller models, by distributing attention more evenly, but both templates converge to similar bias levels in deeper layers across categories and sizes. Overall, model scale remains the strongest factor controlling bias magnitude and stability. Larger models exhibit smoother, more calibrated biases but do not eliminate them-especially in \textit{Risk-} and \textit{Growth}-related prompts-highlighting inherent limitations of general-purpose LLMs when applied to specialized financial tasks.

\noindent \textbf{RQ6: What shared or divergent mechanistic features emerge across model scales and prompt types, and how can these insights inform universal versus context-specific de-biasing strategies for financial LLMs?}
The above analysis reveals that across model scales and prompt types, positional bias consistently centers on a stable set of mid-to-late transformer layers and a small, recurrent group of attention heads that act as core “bias engines.” This universality suggests that fundamental architectural components within Qwen2.5 models are the primary drivers of positional bias regardless of surface-level prompt differences or financial domain, forming natural and effective universal targets for intervention. We found that there are "universal positional bias heads" in all the models as following: Qwen 1.5B: [L16H0, L17H3, L21H9, L21H11, L22H4], Qwen 7B: [L0H7, L17H27, L18H9, L18H25, L23H10] and Qwen 14B: [L25H7, L29H12, L30H7, L30H8, L30H20, L30H38, L32H4, L32H30, L32H33, L35H10, L35H30, L36H21, L36H23], where L and H are Layer and Head numbers. Interestingly, the universal positional bias heads are in the deeper layers (16-23) for smaller models, except L0H7 in 7B model. For 14B model, these universal heads are located in early second-half layers (25-36) but not in the final layers. Notably, smaller models exhibit these bias signals more sparsely but with greater intensity, while larger models show a more distributed yet persistent pattern, reflecting some scale-based smoothing but no eradication of key bias loci.

Simultaneously, important context-dependent divergences modulate how and when bias manifests. Semantic complexity in categories like \textit{Risk}, \textit{Innovation}, and \textit{Growth} amplifies bias strength, especially in smaller models, while simpler categories generate more stable, less biased outputs. Prompt structure influences the timing of bias onset-prompts presenting options before evaluative criteria tend to produce an earlier and more diffuse bias signal-though all converge to the same internal loci deeper in the network. Universal de-biasing should target these stable layers and heads, while context-specific methods like prompt reframing and adaptive monitoring address semantic and phrasing-driven variations, enabling layered, efficient mitigation in financial LLM application.

\section{Conclusion} 
\label{sec:conclusion}

Our results highlight a complex landscape for mitigating positional bias in financial LLMs. Though smaller models may offer new agentic capabilities \cite{belcak2025small}, downsizing increases positional bias, while scaling from 1.5B to 14B parameters substantially reduces it across most financial categories. This aligns with prior findings that larger LLMs reason more reliably. However, the \textit{Risk} category remains stubbornly biased even at a large scale, suggesting domain-specific factors like volatility and exposure anchor the model to positional cues. Additionally, prompt ordering independently influences bias magnitude, with simple reordering causing significant shifts, especially in \textit{Innovation}, emphasizing the need for deterministic prompt templates or randomized prompting with correction.

Most importantly, prompt framing strongly affects bias: \textbf{Conservative}, strong wording amplifies positional heuristics, likely reinforcing primacy bias, while \textbf{Moderate} phrasing promotes balanced evaluation and reduces bias. Our variance analysis attributes about half the bias variation to prompt style, making prompt engineering a powerful, often overlooked control comparable to model scale. For practitioners, addressing positional bias requires a dual strategy: increasing model size to lessen extreme bias and rigorously stress-testing prompt ordering and framing pre-deployment, alongside ongoing monitoring with direction-sensitive metrics to ensure fairness in high-stakes financial applications.

Mechanistic interpretability reveals that positional bias consistently arises and spreads through a stable core of mid-to-late transformer layers and specific attention heads across categories, templates, and model sizes. While larger models reduce bias severity, core “bias engines” persist, unaffected fully by prompt rephrasing. This necessitates a layered mitigation approach—combining universal interventions like targeted head regularization with context-aware strategies such as prompt design, domain-specific fine-tuning, and adaptive bias monitoring—for transparent and reliable LLM use in high-stakes finance.
% Currently, our study is limited to a single model family (Qwen2.5), English-only prompts, and binary pairwise decisions over a FAANG+ universe. Extending this work to other architectures, multilingual contexts, and $k$‑ary decision formats comprises a critical next step. Further, hierarchical modeling is needed to disentangle global versus category-specific effects cleanly. From a mitigation standpoint, promising avenues include representation-level control vectors, activation patching, LoRA-based debiasing adapters, and permutation bagging. Embedding real-time bias sentinels and adaptive prompt re-framing within agentic pipelines will be crucial for responsible, scalable deployment of LLMs in financial domains. Ultimately, our findings confirm that positional bias is scalable but not removable, governed jointly by model size, prompt ordering, and framing style-mandating that prompt design and bias auditing be treated as foundational engineering principles in responsible financial AI.

Currently, our study is limited to Qwen2.5 models, English binary pairwise prompts, and a FAANG+ universe. Future work will broaden to other architectures, languages, and k‑ary decisions, and use hierarchical modeling to separate global from category-specific effects. We also plan to explore mitigations include representation control, activation patching, and prompt re-framing. 

%The following commands are working 
\bibliographystyle{ACM-Reference-Format}
\bibliography{references}

\end{document}